\documentclass[a4paper]{article}
\usepackage{array, diagbox}
\usepackage{enumitem}
\usepackage{INTERSPEECH2020}
\begin{document}

\title{Predicting lexical skills from oral reading with acoustic measures}
\name{Charvi Vitthal\thanks{* Both authors have equal contribution}\sthanks{\hspace{0.2cm}Currently at System Platform Research Laboratories, NEC Corporation, Japan} *, Shreeharsha B S\sthanks{\hspace{0.2cm}Currently an independent researcher} *, Kamini Sabu and Preeti Rao}
\address{Department of Electrical Engineering,\\Indian Institute of Technology Bombay, Mumbai, India}
\email{charvivitthal@gmail.com, harsha11235813@gmail.com, kaminisabu@ee.iitb.ac.in,  prao@ee.iitb.ac.in}

\maketitle

\begin{abstract}
Literacy assessment is an important activity for education administrators across the globe. Typically achieved in a school setting by testing a child’s oral reading, it is intensive in human resources. While automatic speech recognition (ASR) is a potential solution to the problem, it tends to be computationally expensive for hand-held devices apart from needing language- and accent-specific speech for training. In this work, we propose a system to predict the word-decoding skills of a student based on simple acoustic features derived from the recording. We first identify a meaningful categorization of word-decoding skills by analyzing a manually transcribed data set of children’s oral reading recordings. Next the automatic prediction of the category is attempted with the proposed acoustic features. Pause statistics, syllable rate and spectral and intensity dynamics are found to be reliable indicators of specific types of oral reading deficits, providing useful feedback by discriminating the different characteristics of beginning readers. This computationally simple and language-agnostic approach is found to provide a performance close to that obtained using a language dependent ASR that required considerable tuning of its parameters.
\end{abstract}

\vspace{-0.25cm}
\section{Introduction}
Reading aloud has traditionally been an important instructional component in school curricula. Further, oral reading can serve to evaluate, both, a child's word decoding ability and text comprehension \cite{2008miller_RRQ_comprehenprosody}. Good word decoding skills are revealed by the absence, or minimal occurrence, of word-level miscues such as deletions, substitutions and disfluencies. On the other hand, comprehension is indicated by prosodic fluency, which a child typically acquires after word decoding becomes easy enough to free up the necessary cognitive resources \cite{2016breen_FP_comprehension, 2004schwanenflugel_JEP_fluencydecode, 2008miller_RRQ_comprehenprosody}.  Assessment based on oral reading involves having an expert (such as a language teacher) listen to the child reading a chosen text for attributes such as speech rate, correctly uttered words, phrasing and expressiveness. It is thus intensive in human resources. There have been attempts to use ASR to evaluate lexical miscues followed by automatic analyses of the word-level segmentations for prosody evaluation \cite{2017sabu_slate_lets, 2018sabu_sp_prosodyfeatures, 2007black_is_disfluency}. In the reading context, ASR benefits from language models tuned to the intended text. On the other hand, due to the sensitivity to acoustic model training data mismatch, ASR is successful usually when the speaker and environment variability is controlled. In the school scenario, diversity in skill levels and regional accents affect the performance of both the language and acoustic models (LM and AM) which makes the ASR unreliable unless extensively tuned or adapt to the test set.

In this work, we investigate the possibility of using acoustic signal analyses to obtain predictions of certain types of word-decoding disabilities based on potential correlations with acoustic signal features. We first identify the common word-decoding shortcomings by clustering frequencies of specific types of lexical miscues in a corpus of manual transcriptions of a large number of instances of distinct stories read by children across reading skill levels. We investigate the automatic detection of instances of these categories via correlated acoustic signal properties, if any. 
Typically, ASR on a handheld device is achieved online via a server, requiring internet access that may not be available in many rural areas. A device-based approach to reading skill through simple acoustic analyses that flags very poor readers would be a good screening step that provides computational savings over ASR based analyses. \\
\vspace*{-0.02cm}
Traditionally, a student's word decoding skill is defined by the WCPM (words correct per minute) as measured by listening to the read out text. Most research groups working in automatic reading feedback and assessment have focused on improving the performance of an ASR module for reading miscue detection and speech rate measurement \cite{2012black_icassp_tball, 2008li_is_readingontab, 2014cheng_winlpbea_prosody, 2009zechner_SC_speechrater, PearsonVersant}. Research has also addressed reading skill prediction using lexical~\cite{2012black_icassp_tball} or prosodic~\cite{2012mostow_isadept_listen} features or a combination~\cite{2013bolanos_SC_flora} of both. Bolanos et al.~\cite{2013bolanos_SC_flora} consider 2-stage evaluation where the first stage separates the poor readers from good readers, while the second stage discriminates the prosodically good from poor. The features used in both the stages are drawn from the same set of acoustic and lexical features. In the present work, on the other hand, we consider the first stage of separating poor readers using acoustic features alone.\\
\vspace*{0.025cm}
Previous work has exploited fluency metrics such as speaking rate and pause lengths and frequencies to predict non-native adult speaker proficiency in communication settings \cite{2007liscombe_thesis_prosody}. Fontan et al. \cite{2018fontan_is_L2fluency} used low level signal features to estimate speech rate and its regularity in order to predict human ratings of fluency for Japanese learners of French. In Deng et al. \cite{disfluency_feat} measures based on the count and duration of morae (syllable like units) are used to predict fluency levels of recordings of spontaneous speech, however, the features were extracted from a manual transcription. While there are other examples of the correlation of computed acoustic signal features with human expert rated fluency, none of these works attempt to predict specific word decoding attributes from the prosodic analyses. In the present work, we categorize a typical dataset of children's read speech recordings into classes discovered by the unsupervised clustering  of the observed lexical miscues in the manually generated transcriptions. We next investigate acoustic measures to predict the so identified broad categories with a view to develop an automatic system that provides useful descriptions of overall lexical skill. Experimental evaluation of the performance of the system is followed by a critical discussion of the results. 


\section{Data Set and Lexical Analysis}
\subsection{Data Set and Annotation}
\label{annotation}
\begin{figure}[t]
  \centering
  \includegraphics[width=\linewidth]{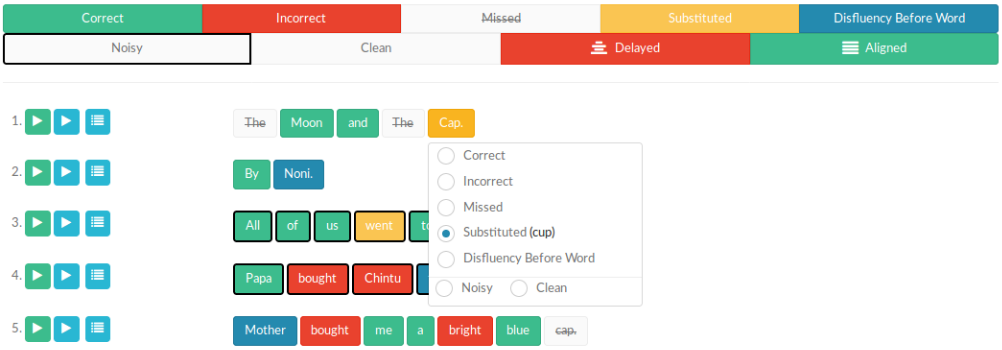}
  \caption{Manual transcription UI with colour-coded lexical miscues marked.}
  \label{fig:transcription_GUI}
\end{figure}

The data set used in the present study comprises oral reading recordings by school children, of the age group 9-13 years in rural schools near Mumbai and Hyderabad, India, learning English as a second language. The story content is tailored to the average 6-8 year old but we see a wide range in proficiency in our data due to the children's acute lack of exposure to the language. A specially created Android app presents the story in video karaoke mode with roughly one sentence per video screen \cite{2016prao_icce_lets}. The words are highlighted in the sequence corresponding to a normal speaking rate, and at the end of the sentence, the video screen switches to the next sentence.  All recordings are made at 16 kHz sampling frequency with a headset mic to minimize background noise (a sample can be found here\footnote{Audio example: https://rebrand.ly/d7byh4k}), and are archived with metadata comprising the child's credentials, story name and date/time of recording. A back-end program facilitates the manual transcription at word level on a web-based interface by presenting pre-segmented sentences of the audio aligned with the intended story text as shown in Figure~\ref{fig:transcription_GUI}. Transcription is carried out by a human transcriber listening to the sentence-level recordings and marking each canonical text word as indicated by the colour codes in Figure~\ref{fig:transcription_GUI} as follows. A stage of validation follows the transcription. 
\begin{itemize}[leftmargin=*]
\item Correct (C) (green) : the word is pronounced correctly.
\vspace{-0.15cm}
\item Missed (M) : the word is skipped, i.e. not uttered at all.
\vspace{-0.15cm}
\item Disfluency (D) (blue) : the word is uttered in incomplete form and/or immediately corrected.
\vspace{-0.15cm}
\item Substituted (S) (yellow) : the word is perceived as a different word(s). Substitution can be by one (S1) or more (Sm) words. The substituted word(s) are keyed in by the transcriber for future use in ASR AM/LM training.
\vspace{-0.15cm}
\item Incorrect (I) (red) : unintelligible speech or gibberish. This is observed to occur typically as a sequence of several syllables with no clear one-to-one correspondence to the expected words.
\end{itemize}

\vspace{-0.1cm}
For our analysis, we have considered 6 distinct stories (each between 10-40 sentences long) read by 208 children to get 1072 audio recordings in all. 
The recording duration ranges from approximately 1 min to 6 min long. 

\vspace*{-0.15cm}
\subsection{Clustering with Lexical Features}
\label{clustering}
\vspace*{-0.15cm}
\begin{figure}[t]
  \centering
    \includegraphics[width=0.75\linewidth,height=100pt]{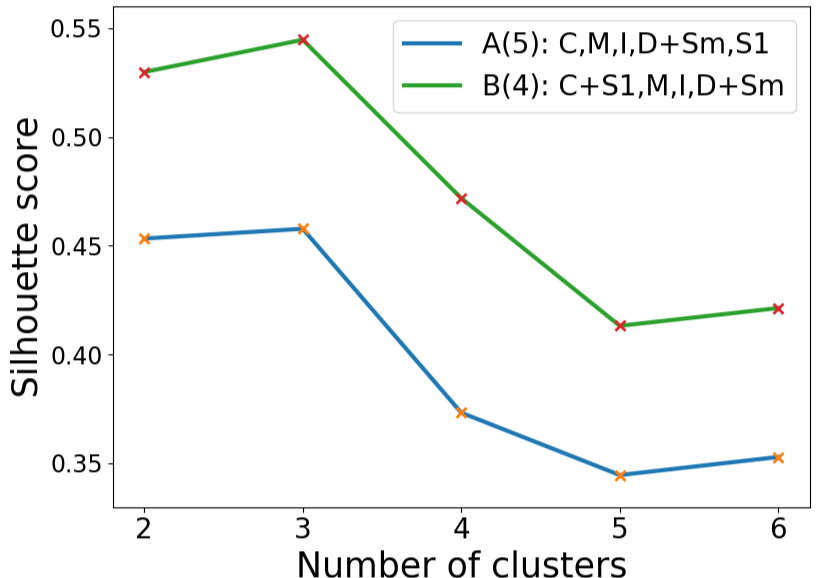}
  \caption{Average silhouette score versus number of clusters for the different feature vectors considered}
  \label{fig:si_score}
\end{figure}
Each audio recording (i.e. one instance of a speaker-story) is represented by the following set of features. From the available manual transcription, we compute the number of each miscue type in the form of a fraction with respect to the total number of words in the given story text. We thus obtain a 5-dimensional feature vector for each recording corresponding to the fractions of words correct (C), substituted (S), missed (M), incorrect (I) and disfluent (D). Since the disfluencies(D) can also be considered as ``substituted by more than one word" (Sm) in the process of self-correction, and the number of occurrences of these two categories is relatively low, we combine these to obtain a 5-dimensional feature vector. We also consider reduced dimension feature vectors via meaningful combinations of the above attributes. We observed that the substitutions by 1 word (S1) are mainly grapheme-to-phoneme pronunciation errors which are very common with Indian language speakers learning English. Therefore, these are not indicative of poor word-decoding skills. We construct a second 4-dimensional feature vector by combining the correct words (C) and substitutions by 1 words (S1).

We apply unsupervised clustering (K-means) to discover possible underlying groupings for different choices of K. We evaluate the quality of the fit by the \textit{silhouette score} that captures the closeness of each sample to its own cluster compared to the closeness to other clusters \cite{silhoutte}. The score (ranging between -1.0 and 1.0) is a measure of how tightly grouped the clusters are, with higher value indicating tighter grouping. Figure~\ref{fig:si_score} shows the silhouette score versus number of clusters. We see that the 4-dimensional vector with the C and S1 combined shows the best clustering.  

\begin{figure}[t]
  \centering
    \includegraphics[width=\linewidth,height=100pt]{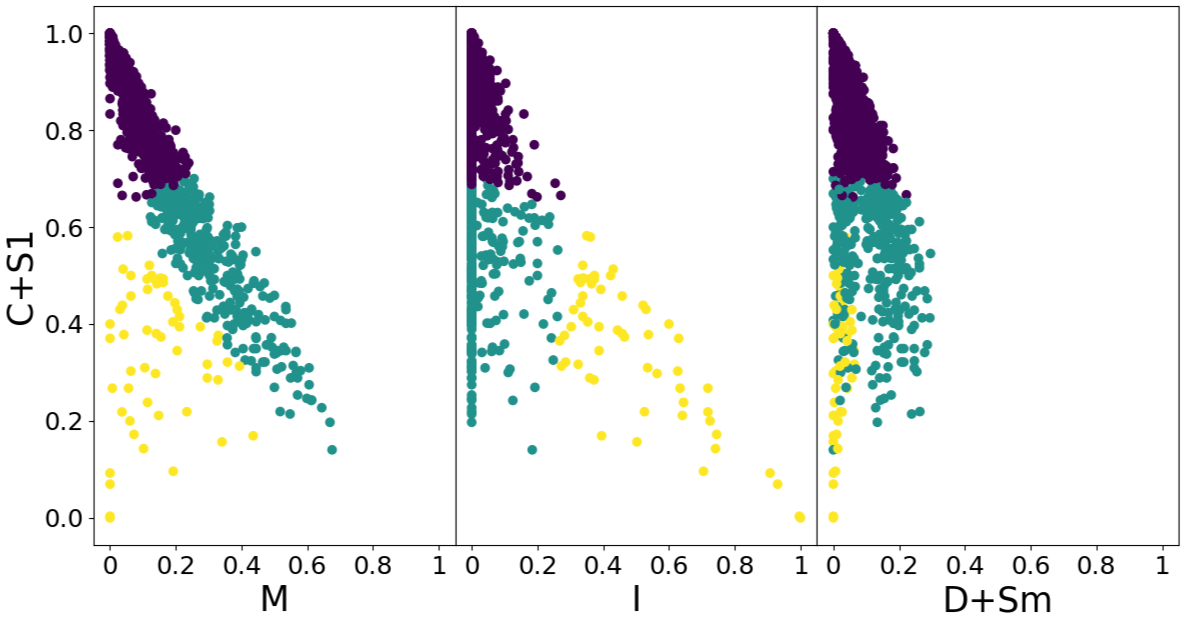}
  \caption{Scatter plots showing 3 clusters obtained using feature vector B(4-dim) of Figure \ref{fig:si_score}. The number of instances shown are: purple (687), blue (329), yellow(56).}
  \label{fig:clusters}
\end{figure}
 \vspace{-0.05cm}
Figure~\ref{fig:clusters} shows the obtained clusters in 2-dimensional space of chosen feature subsets. We can interpret the 3 classes emerging here as follows. Purple colour code corresponds to speech that is predominantly correct (or substituted by a single word). The remaining 2 classes correspond to speakers with a lower proportion of C+S1 words. The latter seems to get further split into cases dominated by gibberish (yellow) versus those dominated by deletions or missing words (blue) speaking a lot of gibberish. This observation provides the interesting insight that children with weak word-decoding skills do not necessarily pause and struggle to decode the individual words but can instead simply skim over the words while speaking a stream of gibberish. We next investigate low level acoustic features, easily computed from a recording by a child, that can provide cues to these 3 distinct lexical skill categories. For fair analysis, we consider a balanced subset of 189 instances drawn from across the 3 clusters.

\vspace{-0.23cm}
\section{Acoustic Features for Classification}
\label{acoustic_feat_classification}
\vspace*{-0.07cm}

We consider features that can be extracted from low-level signal analyses. Features that capture pause behaviour, syllable rate and suprasegmental variations such as intensity and spectral dynamics could help us discriminate proficient speakers from less skilled ones~\cite{2007liscombe_thesis_prosody, 2000cucchiarini_JASA_fluency}. In this work, however, we wish to specifically predict the lexical categories discovered via clustering in the previous section from the low-level signal features. The underlying speech attributes that we take into account, based on our observations and intuition, are the following:
\vspace{-0.15cm}
\begin{enumerate}[leftmargin=*]
    \item Pauses: Silences are obtained from a voice-activity-detection (VAD) module \cite{2017pasad_ncc_vad} that uses energy and harmonicity with temporal smoothness constraints to detect non-speech frames at 10 ms intervals. To exclude plosives, silences exceeding 200 ms are defined as pauses.
    \vspace{-0.15cm}
    \item Syllable-rate: Obtained from sub-band energy peaks (corresponding to the vowels) based syllable nucleus detection~\cite{2007wang_ASLPTran_speechrate, sabu2020sprate}.
    \vspace{-0.15cm}
    \item Dynamics: During speech segments, the short-time spectrum and intensity are expected to vary with changing articulation. Thus heightened dynamics are expected for clearly articulated speech. In the case of unintelligible speech, the dynamics are less prominent due to mumbling at low loudness levels.  
\end{enumerate}
\vspace{-0.10cm}
We present the feature extraction next, followed by classification using these features. 
\vspace*{-0.15cm}
\subsection{Acoustic Feature Extraction}
\vspace*{-0.15cm}
The features corresponding to the broad attributes of speech listed above are computed as follows. A summary of the extracted features appears in Table~\ref{feature_summary}.
\begin{itemize}[leftmargin=*]
\item
\vspace{-0.1cm}
    Pauses: We consider the minimum, maximum, mean and standard deviation (std) of pause duration across the story recording and pause frequency across the video frame intervals which indicate the end of a sentence.
\item
\vspace{-0.1cm}
    Syllable-rate: The number of detected syllable peaks in a video interval is compared with the expected number of syllables in the known text corresponding to the same interval. Next, the mean and standard deviation of this fraction is computed across the video intervals along with the overall articulation rate as the total number of syllables detected divided by the total speech duration of the story recording.
\item 
\vspace{-0.1cm}
   Spectral dynamics (sp-dyn): We compute the spectral centroid over the entire 0 - 8 kHz band from the short-time spectrum every 10 ms frame. Discarding the silence regions based on VAD decisions, in the remaining speech regions in a given video interval, we count the occurrences of the most frequently occurring spectral centroid band and the second most frequently occurring band, c1 and c2 respectively, after dividing the frequency range into bands of 400 Hz width.
The features computed are then:
\vspace{-0.15cm}
    \begin{enumerate}[leftmargin=*]
        \item freq\_distribution ratio (c1/c2):
            Ratio close to one indicates an even energy distribution across the frequency range. Higher ratio indicates single band dominance, characteristic of incorrect speech with repetitive articulation.
        \vspace{-0.15cm}
        \item norm mode count (c1/len(speech in audio)): Finds the proportion of the highest occurring centroid band
        \vspace{-0.15cm}
        \item norm mode variation (std(c1/len(speech in video frame))):
            Measures the deviation, across the recording, of the highest occurring centroid computed every video interval.
    \end{enumerate}
    \vspace{-0.2cm}
    \item Intensity dynamics (int-dyn): The short time intensity contour computed across 10 ms frames exhibits different kinds of dynamics at small and larger (i.e. across syllables) time scales. The intensity contours are more fluctuating in clear articulation and smoother in gibberish. Intensity normalization is performed at the video frame level followed by VAD based silence removal. We compute the following features based on the above observations:
    \vspace{-0.15cm}
    \begin{enumerate}[leftmargin=*]
        \item  Intensity contour smoothness at syllable level:
      For each video interval, we compute the standard deviation of a 300ms moving averaged intensity contour. The mean and standard deviation of this across the recording capture the inter-syllable variation in intensity. 
    
    \item Intensity contour smoothness at micro level: 
    We compute the mean and standard deviation of the contour fluctuations in 10-40 ms time windows. These are expected to measure speaker articulatory variations at a micro level. 
    \end{enumerate}
\end{itemize}

\begin{figure}[t]
\centering
  \includegraphics[width=\linewidth, height = 4cm]{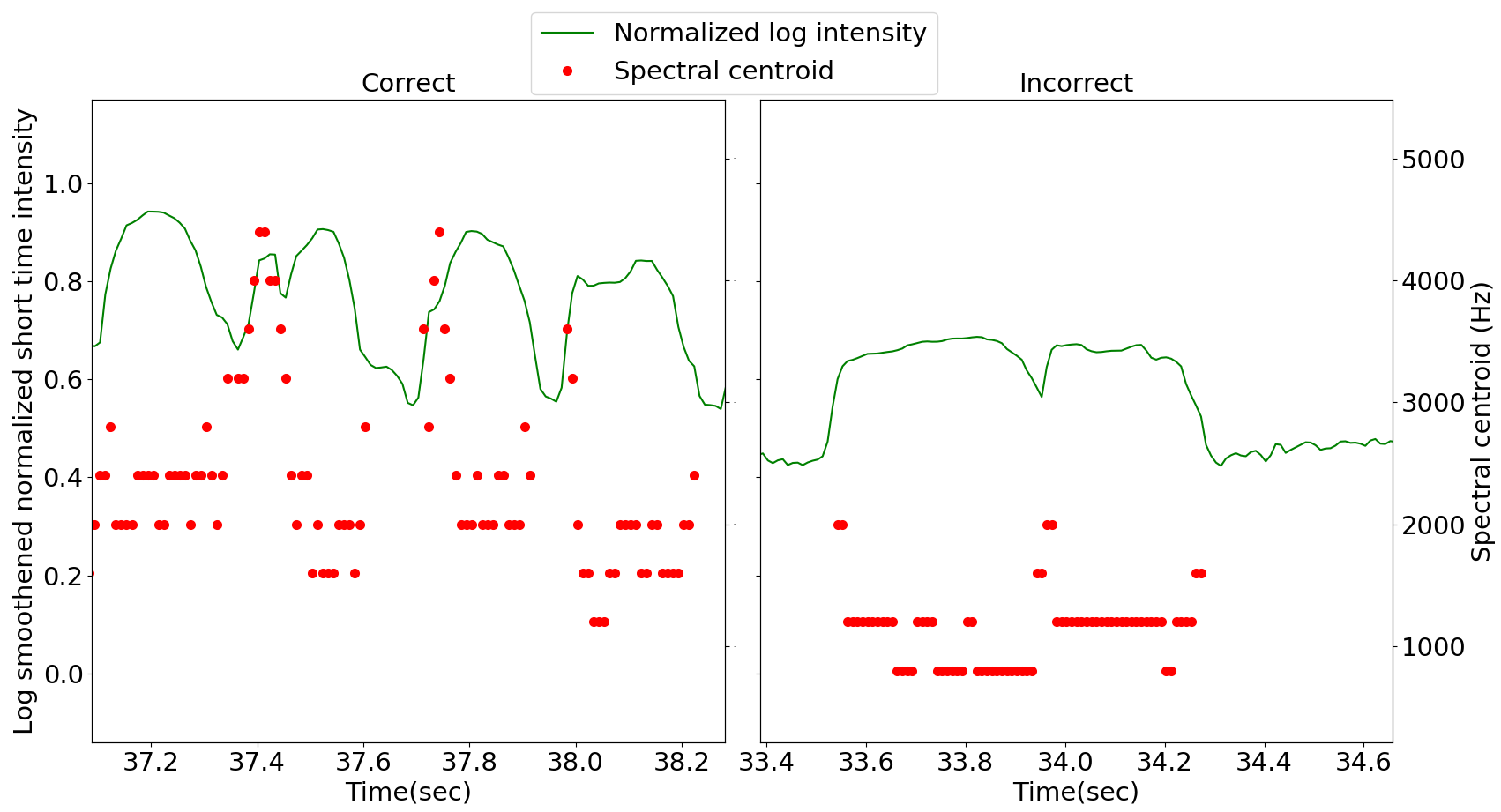}
  \caption{Correct audio. Text:``Branch of an old pipal tree"
  Incorrect audio. Text: ``bran an oll"(branch of an old).} 
  \label{fig:correct}
\end{figure}
\vspace{-0.1cm}
Figure~\ref{fig:correct} shows high dynamic range in intensity contour across syllables for correctly read speech as opposed to the relatively smooth contour for incorrect speech. The red contour in the plot indicates spectral centroid which can be seen to have more flat regions as well as lower values in frequency due to the low vocal effort as well as the limited articulator movements while speaking by children uttering gibberish.
\begin{table}[th]

  \centering
\begin{tabular}{ p{1cm} p{5.75cm}}

    \toprule
    \textbf{Attribute}  & \textbf{Feature}  \\ \hline
    Pause & mean, std, min, max of pause duration, pause freq, \# pauses per video frame \\ \hline
    Syllable rate(SR) & mean, std of relative \# syllables, ratio of std and mean, articulation rate (AR) \\  \hline
    Dynamics & Spectral centroid based (sp-dyn): freq distribution ratio, norm mode count and variation \\
    & Intensity based (int-dyn): Macro, micro-level fluctuations  \\
    \bottomrule
  \end{tabular}
  \caption{Description of extracted acoustic features}
  \label{feature_summary}
\end{table}
  \vspace{-0.2cm}
\vspace{-0.65cm}
\subsection{Classification}
\vspace{-0.10cm}
The acoustic features mentioned in Table \ref{feature_summary} are used for classifying an instance into the classes obtained by lexical clustering of Section \ref{clustering}. A random forest classifier, with 50 trees, is used for this classification. Seven-fold cross validation is performed on the balanced subset of 189 instances specified in Section \ref{clustering} to evaluate the performance. Among these, 70 audios are correct (C$_A$), 63 are missed (M$_A$) and 56 are incorrect (I$_A$) according to the lexical clustering. Next, we compare this resource inexpensive method with an ASR in predicting fluency levels.
\vspace{-0.15cm}
\section{Automatic Speech Recognition for Classification}
\label{classifier_accuracies_ASR}
 An ASR system that uses a DNN-HMM acoustic model which is trained on the speech of children of same age group and similar skill levels, and a tri-gram language model which is trained on the text of the specific story (canonical text) being spoken (along with a garbage model) in Kaldi~\cite{kaldi} is used to predict whether the words spoken in a recording are correct or not. Using the edit distance between the canonical text and the ASR output, each word decoded is labeled as either: Correct(c), Substituted(s), Inserted(i), Deleted(d), which is remapped to the previously found word categories: Correct(C), Incorrect(I), Missed(M). 

Every deleted(d) word corresponds to a missed(M) word, but for the case of substituted(s) words it is not clear whether the word was spoken incorrectly (gibberish) or only slightly mispronounced. To resolve this ambiguity, the posterior probabilities of the 1-best path calculated over the phones from the generated decoded lattice for each test recording are used as a confidence measure relating to how well a hypothesized word was pronounced. Based on this, substituted and inserted words with confidence measures lower than a heuristically chosen threshold are classified as incorrect; those higher as correct.

Mappings between percentages of missed, correct and incorrect words obtained from the ASR output and the ground truth lexical miscue clusters as presented in Sec. 2.2 are derived via the best fit to the full dataset clusters. The mappings are then applied to classify the test recordings into the same three underlying skill categories. We present the classification performances of the two distinct systems next.
\section{Experimental Results and Discussion}

For classification using the acoustic features, we investigate two distinct approaches. First, a single stage classifier is tested with all the extracted features as input vectors per child-story instance. The best performing feature combination and its accuracy are shown in Table~\ref{tab:classifier_accuracies}. 

As the dynamics features are observed to be most efficient in detecting `incorrect' (I$_A$) speech, we also test a 2-stage classifier (P). It is expected to separate I$_A$ from C$_A$ and M$_A$ in the first stage and C$_A$ and M$_A$ in the second stage. In our data, the C$_A$ and I$_A$ classes contain mostly speech as opposed to the M$_A$ class. This motivates another 2-stage classifier (Q) to separate M$_A$ from C$_A$ and I$_A$ first, and then separate C$_A$ and I$_A$. The details of this are present in Table \ref{tab:classifier_accuracies}.
\begin{table}[th]
  \vspace{-0.2cm}
  \centering
\begin{tabular}{l p{3.75cm} c}
    
    
    Classifier  & Feature Combination  & Accuracy \\ 
    configuration &  & \\ \hline
    1-stage & Pause features, SR features, \newline sp-dyn, int-dyn & 65.7 \% \\     \hline 
    2-stage (P) & Stage 1: AR, sp-dyn, int-dyn, \newline\# pauses per video frame \vspace{0.1cm} & 68.3 \% \\
    
     & Stage 2: pause features, \newline SR features & \\  \hline
    2-stage (Q) & Stage 1: pause features, \newline SR features, sp-dyn, int-dyn \vspace{0.1cm} & 64.6 \% \\ 
    
    & Stage 2: sp-dyn, int-dyn, \newline SR features, pause freq   & \\ 
    \bottomrule
  \end{tabular}
  \caption{Classifiers and obtained accuracies in 3-way classification across 189 test instances }
  \label{tab:classifier_accuracies}
  \vspace{-0.15cm}
  
  \centering
\begin{tabular}{|c|c|c|c|}
\hline
   \diagbox[width = 3cm, height = 5ex]{Actual}{Predicted}  & \textbf{C$_A$} & \textbf{M$_A$} & \textbf{I$_A$} \\ \hline
    \textbf{C$_A$}& 51 & 7 & 12\\    \hline
\textbf{M$_A$} & 15 & 35 & 13\\    \hline
 \textbf{I$_A$}& 7 & 6 & 43\\    \hline
 
  \end{tabular}
  \caption{Confusion matrix for the best classifier in Table 2} 
  \label{tab:confusion_matrix}
\end{table}
Table \ref{tab:confusion_matrix} shows the confusion matrix corresponding to the best classification scheme (P) of Table \ref{tab:classifier_accuracies}. 
The misclassifications are found to be more in the missed class compared to the others; this is because whenever the students in the missed class spoke, they were correct (largely) or incorrect. We can say that the missed class bridges the gap between correct and incorrect. This can also be observed from cluster plots in Figure \ref{fig:clusters}. We have also observed that the correct class is more confined whereas the other two classes display varying amounts of their respective characteristics. So, we expect more confusion in classifying these. The dynamics features chosen, sp-dyn and int-dyn, are designed to pick the incorrect (gibberish) speaking students from the others, improving the classification of incorrect class. This motivated the two-stage classification (P). On the other hand, we were unable to find similar tailored features for the missed class, apart from the pauses, which are adversely affected by different kinds of noise, ranging from mic noise to background talkers. We also note, similar to Deng et al. \cite{disfluency_feat}, that our pause based features were the least important according to the RF classifier while predicting the fluency categories, although the noise is also a contributing factor.
This explains the comparatively lower accuracies for the missed class in Table 3 

The ASR performance was carefully tuned by varying the language model weights and re-scoring options to get the best results, presented in Table 4. 
The overall classification accuracy obtained is 72.5\%. The ASR errors are related with not being able to identify incorrect words, and noise, background talkers being picked up as substituted words drawn from the garbage model. On examining many of these chunks, it was found that few of the words (such as 'Washington' and 'castle') that were determined as incorrect by the transcriber were being found as substituted/correct by the ASR with a high degree of confidence. Due to the constraints created by the language model trained on the story text, a word that is rare enough and poorly spoken could be classified as correct because of how rare that specific combination of phones is. These events happen with a garbage model and other OOV (out of vocabulary) words as well. This indicates that the ASR might not be as adept at picking up gibberish words because even slightly similar acoustic characteristics of the test word compared with the expected word gives rise to high confidence scores.
\vspace{-0.25cm}
\begin{table}[th]
  \centering
\begin{tabular}{|c|c|c|c|}
\hline
   \diagbox[width = 3cm, height = 5ex]{Actual}{Predicted}  & \textbf{C$_A$} & \textbf{M$_A$} & \textbf{I$_A$} \\ \hline
\textbf{C$_A$}& 51 & 10 & 9\\    \hline
\textbf{M$_A$} & 11 & 40 & 12\\    \hline
 \textbf{I$_A$}& 6 & 4 & 46\\    \hline
  \end{tabular}
  \caption{Confusion matrix of the ASR classification of Section \ref{classifier_accuracies_ASR}}
\end{table}
\vspace{-0.8cm}
\section{Conclusion}
\vspace*{-0.15cm}
Based on the fact that beginner (second-language) readers come with diverse skill levels in the word-decoding aspect of oral reading, we attempted to characterize the behaviour of the population represented in our children's reading data set. The best underlying clusters in lexical miscues space turned out to correspond to good readers and two types of lower-proficiency readers. It emerged that poor word decoders can not only skip words they cannot recognize, but also resort to speaking gibberish (unintelligible stream unconnected to the text). We proposed acoustic signal features to discriminate the incorrect (unintelligible) speech from correct speech, and overall, achieve the categorization of speakers into the 3 classes. We compare the performance of the proposed system with that of a full-blown ASR-based system. The ASR based system required the tuning of parameters to obtain the best classification result of 72.5\%. While this is slightly superior compared to the accuracy achieved by acoustic feature classification, we note that the latter system is language agnostic requiring knowledge only of the total number of syllables in the canonical text. The ASR system performance is highly dependent on the AM-LM weighting and thresholds used. The acoustic features are seen to be more general in their applicability and easily deployable with lower computational complexity. Future work will target a larger data set and a closer study to improve the acoustic features further while considering their robustness to background noise.

\bibliographystyle{IEEEtran}

\bibliography{template}


\end{document}